\documentclass[aps,prd,twocolumn, groupedaddress,amssymb,eqsecnum,showpacs,8pt,showkeys,epsfig,nofootinbib]{revtex4}
\usepackage{epsfig}
\usepackage{graphics}
\usepackage{bm}
\usepackage{color}
\usepackage{dcolumn}   
\usepackage[spanish,english]{babel}
\usepackage{bm}     
\usepackage{bbm}       
\usepackage{amssymb}  
\usepackage{amsmath}
\usepackage{latexsym}
\usepackage{float}
\usepackage{ifthen}
\usepackage{caption,subfig}
\usepackage{enumerate}
\usepackage{url}
\usepackage{caption,subfig}
\usepackage{amsopn}
\usepackage{hyperref}

\def \lleq {\lower0.9ex\hbox{ $\buildrel < \over \sim$} ~}
\def \ggeq {\lower0.9ex\hbox{ $\buildrel > \over \sim$} ~}

\def \beq  {\begin{equation}}
\def \eeq  {\end{equation}}
\def \ber  {\begin{eqnarray}}
\def \eer  {\end{eqnarray}}

\begin{document}
\newcommand{\newc}{\newcommand}
\newc{\ben}{\begin{eqnarray}}
\newc{\een}{\end{eqnarray}}
\newc{\be}{\begin{equation}}
\newc{\ee}{\end{equation}}
\newc{\ba}{\begin{eqnarray}}
\newc{\ea}{\end{eqnarray}}
\newc{\bea}{\begin{eqnarray*}}
\newc{\eea}{\end{eqnarray*}}
\newc{\D}{\partial}
\newc{\na}{\nabla}
\newc{\ie}{{\it i.e.} }
\newc{\eg}{{\it e.g.} }
\newc{\etc}{{\it etc.} }
\newc{\etal}{{\it et al.}}
\newcommand{\nn}{\nonumber}
\newc{\ra}{\rightarrow}
\newc{\lra}{\leftrightarrow}
\newc{\lsim}{\buildrel{<}\over{\sim}}
\newc{\gsim}{\buildrel{>}\over{\sim}}
\newc{\G}{\Gamma}
\title{Exact Classical and Quantum Solutions for a Covariant Oscillator Near the Black Hole Horizon in Stueckelberg-Horwitz-Piron Theory }
 
 \author{Davood Momeni$^{1,2}$ }

\affiliation{\it $^1$ Department of Physics, College of Science, Sultan Qaboos University,\\P.O. Box 36, ,AL-Khodh 123  Muscat, Sultanate of Oman \\$^2$Tomsk State Pedagogical University, TSPU, 634061 Tomsk, Russia  
}
 
\date{\today}

\begin{abstract} 
We found  exact solutions for canonical  classical and quantum dynamics for general relativity in   Horwitz general covarience theory. These solutions  can be obtained by solving the generalized geodesic equation  and  Schr\"{o}dinger-Stueckelberg -Horwitz-Piron (SHP) wave  equation for a simple harmonic oscilator in the background of a two dimensional dilaton black hole   spacetime metric. We proved the existence of an orthonormal basis of eigenfunctions for generalized   wave equation. This basis functions form an orthogonanl and normalized (orthonormal) basis  for an appropriate  Hilbert space. The energy spectrum  has a mixed spectrum with one conserved momentum $p$ according to  a quantum number $n$. To  find the ground state energy we used a variational method with appropriate boundary conditions. A set of mode decomposed  wave functions and calculated for the  Stueckelberg-Schrodinger equation  on a general five dimensional blackhole spacetime in Hamilton gauge. 
\end{abstract}

\pacs{ 03.30.+p, 03.65.-w, 04.20 Cr, 04.60 Ds, 04.90.+e}
\keywords{General relativity, quantum theory on curved space, canonical quantum gravity }

\maketitle
\section{Introduction}
General relativity (GR) is considered  so far as the best classical gauge theory for gravity. It has passed  many observational tests locally and generally (for a complete description of gravity as gauge theory see \cite{Moshe}). A fundamental basis to build any modification to GR must preserve the equivalence principle where two frames can be related to each other just by infinitesimal transformations. Building a relativistic quantum mechanics in a canonical formalism has an old history, started by the works of  Stueckelberg\cite{Stueckelberg} , who defined the covariant canonaical classical and quantum theory for one body systems, with its covarint Stueckelberg-Schr\"odinger wave equation.   The theory was generalized to be applicable to many body systems by Horwitz and Piron\cite{SHP} (which we will refer to as SHP theory) and developed more fully by Horwitz \cite{SHP}].
\par Field theories on curved spaces have been studied, for example, by
Birrell and Davies \cite{Birrel}, where there are a list of difficulties. 
In SHP theory just by simple replacement of the classical Poisson brackets with Dirac commutators one can construct  suitable basis functions for an appropriate Hilbert space. This last step also was succesfully ended and so far we have a fully relativistic  wave equation. It was not so easy and clear to generlize all the above ideas from the flat  pseudo Minkowski  spacetime to the arbitrary curved spacetime (see for example \cite{Birrel} where we can find a list of difficulties regarding definition of vacuum state, renormalization and regularization , etc. ).  We are still very far from a "good" and "well defined" quantum gravity and still we believe "semiclasical quantum gravity" or what is called "quantum effects on the curved background" is the best strategy handle the  problem \cite{Mukhanov}. Very recently Horwitz revisited SHP theory and generalized it to the general Lorentzian curved manifold  \cite{arXiv:1810.09248}.  Horwitz's theory  is  a significant work in the field, because he only used local coordinate transformations as a general coordinates transformation betwen frames. In his derivation he followed the original procedure of Einstein to built his ground breaking GR theory. The point is that he built both classical and quantum version of dynamics of a test particle subjected to the  potential . In Horwitz's derivation of dynamical equations  the potential appears as a spacetime mass distribution. He extended his ideas to  electromagnetism using the same method as  Yang's pioneering development of electromagnetism  as a $U(1)$ fiber bundle in his theory\cite{Yang}. In Horwitz's canonical dynamics for electromagnetic fields, $U(1)$ field emerged  as a gauge field on the quantum mechanical Hilbert space. In this work I study exact solutions for classical and quantum dynamical equations in his theory for a simple harmonic oscilator as an external force field. 
\section{Classical solutions }
As  pointed out in the SHP theory,the classical dynamics of a test particle in a potentail field $V(x)$ can be derived using a local coordinate transforamtion from  $\xi^{\mu}\to x^{\mu}$  transformation :
\begin{eqnarray}
&&H=\frac{M}{2}g_{\mu\nu}\dot{x}^{\mu}\dot{x}^{\nu}+V(x)\label{H}.
\end{eqnarray}
Here $V(x)$ on a manifold can be  the potential at point $\xi$ as dual coordinate to $x$, and $g_{\mu\nu}$ is  the spacetime metric. In SHP theory, two aspects of time are distinguished:  the chronological or historical time $\tau$, and the coordinate time $t$.  The latter is a dynamical variable in the manifold, while the former parameterizes motion of an event in the manifold.
We consider only $\tau$ independent metrics of spacetime i.e. those  static spacetime metrics with $\frac{\partial{g}_{\mu\nu}}{\partial\tau}=0$. The  exterior region is supposed to be an asymptotically flat Einsteinian manifild , i.e., a solution to the Einstein gravitarional field equation can be studied. Here our aim is to study exact solutions for dynamical field equations derived from the Hamiltonian (\ref{H}). It is easy to derive a modified  geodesics equation using least action principle and it yields to the following set of coupled nonlinear  second order differential equations for the coordinates $x^{\mu}$:
\begin{eqnarray}
&&\ddot{x}^{\sigma}+\Gamma^{\sigma}_{\lambda\gamma}\dot{x}^{\lambda}\dot{x}^{\gamma}=-\frac{1}{M}g^{\sigma\lambda}\frac{\partial V}{\partial x^{\lambda}}\label{geodesic}.
\end{eqnarray}
Note that Eq. (\ref{geodesic}) gives  orbits different from  the classical geodesic equations. We consider a simple harmonic oscillator near the horizon of a two dimensional black hole, possessing an asymptotically flat metric defined by the line element $ds^2=g_{\mu\nu}dx^{\mu}dx^{\nu}$:
\begin{eqnarray}
&&ds^2=-f(r)dt^2+\frac{dr^2}{f(r)}\label{g}.
\end{eqnarray}
It is adequate to use null coordinates $(x_{+},x_{-})$ where the metric (\ref{g}) converts to the following conformally flat metric:
\begin{eqnarray}
&&ds^2=f(x_{+},x_{-})dx_{+}dx_{-}\label{g2}.
\end{eqnarray}
The reason to use this low dimensional gravity toy model is first of all its simplicity, secondly this gravitational model can be derived from dimensional reduction of the  Einstein-Hilbert action from higher dimensional spacetimes as well as the near horizon geometry of four dimensional blackholes. In metric (\ref{g2}), the pair of null coordinates are defined as
\begin{eqnarray}
&&dx_{\pm}=\frac{dr}{|f(r)|}\mp dt\label{x12}.
\end{eqnarray}
Note that $f(r)>0$ for exterior region of the blackhole for $r>r_h$, where $r_h$ is location of horizon of blackhole. We are working on the exterior region of blackhole spcetime metric. In our toy model (\ref{g2}) , we consider $V(r)=V(x_{+},x_{-})$ as a simple harmonic oscilator as an external force to modify geodesic equation equation in the curved background $g_{\mu\nu}$. By integrating Eq. (\ref{x12}) we find:
\begin{eqnarray}
&&x_{+}=\int\frac{dr}{f(r)}- t+c_{+}\label{x1},\\&&
x_{-}=\int\frac{dr}{f(r)}+ t+c_{-}\label{x2}.
\end{eqnarray}
here $(c_{+},c_{-})$ are integration constants. If two dimensional (2D) gravity posses asymptotic flat regime, i.e, $r\to\infty$, $f(r)\to 1$ then 
\begin{eqnarray}
&&x_{+}\approx r- t+c_{+}\label{x11},\\&&
x_{-}\approx r+ t+c_{-}\label{x22}.
\end{eqnarray}
In the asymptotic geometry of  (\ref{g2}), we take the particle mass squared to be the "on shell" value $M^2$.  We mention here that $c_{\pm}$ play the roles of a zero point energy(chemical potential) for the system. The reason is that when the system remains near asymptotic limit , 
\begin{eqnarray}
&&V(r)\approx \frac{1}{2}k(\frac{x_{+}+x_{-}-c_0}{2})^2\label{v}
\end{eqnarray}
where $k$ is the spring constant and $c_0$ is the location of the equlibrium point, i.e.,  $x^{eq}_{+}+x^{eq}_{-}=c_0$. This equilibrium point $(x^{eq}_{+},x^{eq}_{-})$  is obtained for $\frac{\partial V}{\partial x_{\pm}}|_{(x^{eq}_{+},x^{eq}_{-})}=0$.\par
Note that now the metric (\ref{g2}) looks like just a conforaml flat metric:
\begin{eqnarray}
&&ds^2=f(x^A,x^B)\eta_{AB}dx^Adx^B\label{g3}.
\end{eqnarray}
here $\eta_{AB}=diag(-1,+1)$ and $x^A=(x_{+},x_{-})$. Solving (\ref{geodesic}) for metric (\ref{g3}) needs all non zero Christoffel symbols :
\begin{eqnarray}
&&\Gamma^{A}_{BC}=\frac{1}{2}g^{AD}\big(g_{BD,C}+g_{CD,B}-g_{BC,D}
\big).
\end{eqnarray}	
here $g_{BD,C}\equiv \frac{\partial g_{BD}}{\partial x^C}$ and etc. We use a formula  relating two  conformal metrics; if $\tilde{g}=e^{2\Omega}g$, 
\begin{eqnarray}
&&\tilde \Gamma^{k}_{ij}=\Gamma^{k}_{ij}+\Big(\delta_{i}^k\frac{\partial \Omega}{\partial x^j}+\delta_{j}^k\frac{\partial \Omega}{\partial x^i}-g_{ij}g^{kl}\frac{\partial \Omega}{\partial x^l}
	\Big).
\end{eqnarray}

where in our case $\Omega=\frac{1}{2}\ln f$ and $\Gamma^{k}_{ij}=0$. The potential function $V$ is given in Eq. (\ref{v}) the gradient components are :
\begin{eqnarray}&&
\frac{\partial V}{\partial x^A}=\sqrt{\frac{kV}{2}}(1,1).
\end{eqnarray}
Finally we obtain geodesic equations (\ref{geodesic}) as a pair of coupled nonlinear differential equations:
\begin{widetext}
\begin{eqnarray}
&&\ddot{x}_{\pm}+\frac{1}{2}\dot{x}_{\pm}\frac{d(\ln f)}{dt}-\dot{x}_{+}\dot{x}_{-}
\frac{\partial \ln f}{\partial x_{\pm}}+\frac{1}{Mf}\sqrt{2kV}=0.
\label{geodesic2}.
\end{eqnarray}
\end{widetext}
The asymptotic solution for Eq. (\ref{geodesic2}) is given as follows:
\begin{eqnarray}
&&x_{+}-x_{-}\approx v_0\tau+w_0,\\&&
t(\tau)\approx \frac{\tau}{\tau_0}+t_0.
\end{eqnarray}
The goal is to find exact solution for Eq. (\ref{geodesic2}) with potential (\ref{v}). For 2D blackhole a possible metric function in $f(r)=\lambda^2r^2-\frac{m}{\lambda}$. We can rewrite the metric function in terms of the horizon radius for which $f(r_h) =0$ as follows:
\begin{eqnarray}
&&f(x_{+},x_{-})= \frac{2m/\lambda}{e^{-y}-1},\\&&y\equiv 2r_h\lambda^2(\sum_{a=\pm} x_a-c_0)\label{f}.
\end{eqnarray}
It is remarkable that Eq. (\ref{f}) looks like the Bose-Einstein distribution functions of density of states with energy $E$. We  represent the potential $V$ as $V=\frac{1}{2}kr_h^2\coth^2(y/2)$. If we define a new dependent variable $w\equiv \sum_{a=\pm} x_a$ and  geodesic equation (\ref{geodesic2}) reduces to :
\begin{eqnarray}
&&\ddot{w}+r_h\lambda^2\frac{\dot{w}^2}{1-e^{2r_h\lambda^2(w-c_0)}}=0\label{w}.
\end{eqnarray}
If we substitute $p=\dot{w}$, we obtain (one is given implicitly) two types of solutions for $w(\tau)$:
\begin{widetext}
\begin{eqnarray}
&&w(\tau):
\left\{
\begin{array}{ll}
=\mbox{constant}  & \mbox{if } \dot{w}=0 \\
e^{-y} \sqrt{1-e^{2 y}}+ \sin ^{-1}(e^y)=c_2-(\frac{
	\lambda ^2 r_h e^{c_0 \lambda ^2r_h}}{c_1})\tau
&  \mbox{if } \dot{w}\neq 0
\end{array}
\right.
\end{eqnarray}
\end{widetext}
where $w=c_{0}+\frac{y}{\lambda ^2 r_h} $.

Finally we find a classical  geodesic solution for SHP theory on a 2D toy model for BH . It is adequate add a few words about the importanvce of the classical solution obtained here. The above solution for $w(\tau)$ indentifies the classical path of $x_{+}+x_{-}$  as a function of the chronological or historical time $\tau$. At past when $\tau\to-\infty$  the right hand side of the solution for $w(\tau)$ tends to $-\infty$, a possible solution for $y$ can be obtained if we consider the asymptotic limit $y \to-\infty$; in this asymptotic regime, we obtain:
\begin{eqnarray}
&&w(\tau)\approx
\-(\lambda ^2 r_h)^{-1}\ln(\frac{\tau}{\tau_{-\infty}})+c_0
\end{eqnarray}
Plugging this asymptotic solution in the metric form given in the equation (\ref{g2}) we will find de Sitter metric in two dimensions. The exisence of the de Sitter solution is an important element to describe early cosmology in favor of the inflation scenario. We mention here that the metric given in the equation (\ref{g2}) can be considered as an effective dilatonic field and the metric field equations can be treated as equations of motion for such dilatonic field. Consequently our classical solution supports the idea of classical fields as effective fields decribe early eraly(pre early) inflation. Further investigations can be proceed to find perturbations of metric and thermal spectrum of the matter contents using this classical solution. Furthermore, this asymptotic solution can be used to build the appropriate forms for the Green function. It will provide the  chronological time propagators and using those propagators we can study the causal structeres of the spacetime metric at early universe.

\section{Quantum mechanics via Schr\"{o}dinger-Stueckelberg -Horwitz-Piron wave equation on 2D blackhole background }
In the previous section we found the classical trajectory of a test particle subjected to a harmonic oscilator classical force on a 2D blackhole background. There is a well formulated quantum mechnaical wave equation for quantum dynamics on a curved general relativity background well formulated in \cite{ arXiv:1810.09248}:
\begin{eqnarray}
&&i\hbar\frac{\partial}{\partial\tau}\Psi_{\tau}(x^\mu)=\hat{H}\Psi_{\tau}(x^\mu)\label{waveeq1}
\end{eqnarray}
where $\hat{H}$ is quantum mechanical operator form of Eq. (\ref{H}). It is defined on a Hilbert space with scalar product :
\begin{eqnarray}
&&(\psi,\chi)=\int d^4x\sqrt{-g} \Psi^{*}_{\tau}(x^\mu)\chi_{\tau}(x^\mu)
\end{eqnarray}
where the volume element is written as a normal scalar and $^{*}$ denotes complex conjugate. For a quantum mechanical Hamiltonian $\hat{H}$ we follow the convention of indexes given in ref.  \cite{arXiv:1810.09248}. We define the Hamiltonian as:
\begin{eqnarray}
&&\hat{H}=\frac{1}{2M\sqrt{-g}}p_{\mu}g^{\mu\nu}p_{\nu}+V(x)\label{H1}.
\end{eqnarray}
The potential is given as a quantum harmonic oscillator i.e, Eq. (\ref{v}). On the background given in the spacetime metric (\ref{g3}) we find $\sqrt{-g}=f/2$, using the momentum operator given by $p_{\mu}=-i\hbar\partial_{A},x^A=(x_{+},x_{-})$; supposing that $\Psi_{\tau}=\Psi_{\tau}(x_{+},x_{-})$, we  reduce the wave equation Eq. (\ref{waveeq1}) to the following time independent schr\"{o}dinger-Stueckelberg -Horwitz-Piron wave equation in the null coordinates :
\begin{widetext}
\begin{eqnarray}
&&-\frac{\hbar^2}{2Mf^2}\Big(2\partial_{+}\partial_{-}\phi(x_{+},x_{-})-\Big(\partial_{+}(\ln f)\partial_{-}\phi(x_{+},x_{-})+\partial_{-}(\ln f)\partial_{+}\phi(x_{+},x_{-})
\Big)
\Big)\\&&\nonumber+V(x_{+},x_{-})\phi(x_{+},x_{-})=E\phi(x_{+},x_{-})
\label{waveeq2}
\end{eqnarray}
\end{widetext}
where we use seperation of the variables as $\Psi_{\tau}(x_{+},x_{-})=e^{-i\frac{E\tau}{\hbar}}\phi(x_{+},x_{-})$. Note that $\partial_{\pm}f(x_{+},x_{-})=\frac{2r_h\lambda^2}{1-e^{y}}$ where $y$ is defined in 
Eq. (\ref{f})  at temperature $T$. We then obtain the following  partial differential equation for the time independent wave function $\phi\equiv\phi(x_{+},x_{-})$:
\begin{widetext}
	\begin{eqnarray}
	&&-\frac{\hbar^2\lambda^2}{2mM}\Big(e^{-y}-1
	\Big)\Big[\partial_{+}\partial_{-}-\frac{r_h\lambda^2}{1-e^{y}}
(\partial_{+}+\partial_{-})	\Big]\phi+\Big[\frac{1}{2}kr_h^2\coth^2(\frac{y}{2})\Big]\phi=E\phi
	\label{waveeq3}
	\end{eqnarray}
\end{widetext}
If we use new coordinates $\xi\equiv x_{+}+x_{-},\zeta\equiv x_{+}-x_{-}$ and using separation of variables; (note that $\zeta$ is a cyclic variable, the  momentum $p$ corresponding to $\xi$ is a conserved quantity), $\phi(x_{+},x_{-})=e^{\frac{ip\zeta}{\hbar}}F(\xi)$ and
we end up with the following second order differential equation:
\begin{widetext}
	\begin{eqnarray}
	&&\frac{d^2F(y)}{dy^2}-\frac{\bar{r}^2}{1-e^y}\frac{dF(y)}{dy}+\Big[
\bar{k}_{p,n}^2-\bar{m}^2\frac{e^y\coth^2y}{1-e^y}	\Big]F(y)=0
	\label{waveeq4}
	\end{eqnarray}
\end{widetext}
here $\bar{r}^2=r_h\lambda^2,\bar{m}^2=\frac{mMkr_h}{2\hbar^2\lambda^4},\bar{k}_{p,n}^2=\frac{mME}{r_h\hbar^2\lambda^4}+\frac{p^2}{2r_h\lambda^2\hbar^2}$.
An appropriate set for  boundary conditions is :

	\begin{eqnarray}\label{bc}
	&&\Psi_{\eta}(y,\zeta)=
	\left\{
	\begin{array}{ll}
	0  & \mbox{if } y\to 0 \\
	0
	&  \mbox{if } y\to-\infty
	\end{array}
	\right.
	\end{eqnarray}

After a linear transformation
 in the form $F(y)=\chi(y)\exp{(e^{-y}-1)^{-\frac{\bar{r}^2}{2}}} $ the resulting differential equation for $\chi(y)=\chi$ is 
\begin{widetext}
	\begin{eqnarray}
	&&\chi''+\Big(\bar{k}_{p,n}^2-\frac{\bar{r}^4}{4(1-e^y)^2}-\bar{m}^2\frac{e^y\coth^2y}{1-e^y}+\frac{\bar{r}^2e^y}{2(1-e^y)^2}
	\Big)\chi=0
	\label{chi-eq}
	\end{eqnarray}
\end{widetext}
The differential equation (\ref{chi-eq}) has two asymptotic solutions given by 

\begin{eqnarray}
&&\chi(y)=
\left\{
\begin{array}{ll}
Ay\log{y}& \mbox{if } y\to 0 \\
B\sin(\bar{k}_{p,n} y+\theta_0)	&  \mbox{if } y\to-\infty
\end{array}
\right.
\end{eqnarray}
respectively. Thus, the general solution of equation (\ref{chi-eq}) assumes the form
\begin{eqnarray}
&&\chi(y)=
y\log{y}\sin(\bar{k}_{p,n} y+\theta_0)	g(y)\label{ansatz1}
\end{eqnarray}
By substituting the ansatz (\ref{ansatz1}) in Eq. (\ref{chi-eq}), it is straightforward to show that the auxiliary function $g(y)$ satisfies another second order differential equation which cannot be solved analytically.
\par 
In the absence of such analytic solutions , we   investigate the differential equation Eq. (\ref{waveeq4}) using Sturm-Liouville theory. The energy spectrum for Eq. (\ref{chi-eq}) is a  mixed spectrum given by 
\begin{eqnarray}
&&E_{p,n}=\frac{r_h\hbar^2\lambda^4}{mM}(\bar{k}_{p,n}^2-\frac{p^2}{2r_h\lambda^2\hbar^2}
)\label{E}
\end{eqnarray}
We can  multiply (\ref{waveeq4})   by a quantity that converts it into self-adjoint form.The Sturm-Liouville eigenvalue problem of Eq. (\ref{waveeq4}) becomes:
\begin{widetext}
\begin{eqnarray}
&&\frac{d}{dy}\Big[(e^{-y}-1)^{-\bar{r}^2}F_{p,n}'	\Big]+\Big[
\bar{k}_{p,n}^2(e^{-y}-1)^{-\bar{r}^2}-\bar{m}^2(e^y\coth^2y)(e^{-y}-1)^{-1-\bar{r}^2}	\Big]F_{p,n}(y)
=0\label{SL}
\end{eqnarray}
\end{widetext}
It is then straightforward to show that the equation  (\ref{SL}) satisfies the self-adjoint condition. The solutions satisfy
\begin{widetext}
	\begin{eqnarray}
	&&\int_{-\infty}^{0}F_{p,n}(y)F_{p,m}(y)(e^{-y}-1)^{-\bar{r}^2}dy=
	\left\{
	\begin{array}{ll}
	0  & \mbox{if } m\neq n \\
	N_{p,n}
	&  \mbox{if } m=n
	\end{array}
	\right.\label{orthogonalization }
	\end{eqnarray}
\end{widetext}
The wave function of a general quantum state is given by the following superposition expression:
\begin{widetext}
\begin{eqnarray}
	&&\Psi_{\tau}(x_{+},x_{-})=\sum_{n=0}^{\infty}\int_{-\infty}^{\infty}dp	 e^{\frac{i}{\hbar}(-E_{p,n}\tau+p( x_{+}-x_{-}))}  C_{p,n}F_{p,n}\Big( 2r_h\lambda^2( x_{+}+x_{-}-c_0)\Big),
\end{eqnarray}
\end{widetext}
here $C_{p,n}$ is the Fourier coefficient for the series and can  be obtained using initial conditions $\Psi_{\tau}(x_{+},x_{-})|_{\tau=0}$. This quantum mechanical wave function deserves to be investigated in details. First of all , as we mentioned the form of the wave function can be fixed if we know enough initial condition about the shape of the wave function on an initial chronological time i.e., $\Psi_{\tau=0}(x_{+},x_{-})$. This initial wave packet specfies the initial state of the quantum state on two dimensional background. The probability of finding a state with momentum $p$ and on  the level $n$ is given by $|C_{p,n}|^2$. This probability amplitude can be obtained using the orthogonalization condition (\ref{orthogonalization }) as follows:
\begin{widetext}
	\begin{eqnarray}
	&& C_{p,n}=\int_{-\infty}^{\infty}\Psi_{\tau=0}(y)e^{-\frac{i}{\hbar}p( x_{+}-x_{-})} F_{p,n}(y)(e^{-y}-1)^{-\bar{r}^2}dy,
	\end{eqnarray}
\end{widetext}
	Furthermore, we can use the above amplitude to compute different types of probabilties, for example if we are looking for the probability of finding the system in ground state $n=0$ with large momenta $p_{UV}$ is :
\begin{widetext}
	\begin{eqnarray}
	&& C_{p_{UV}}=\int_{-\infty}^{\infty}\Psi_{\tau=0}(y)e^{-\frac{i}{\hbar}p_{UV}( x_{+}-x_{-})}F^{0}_{p_{UV}}(y)(e^{-y}-1)^{-\bar{r}^2}dy,
	\end{eqnarray}
\end{widetext}	
The quantum mechanical amplitude on the above line can be used to find the most probable large momentum state which can be used to explain high energies excitations in the vicinity of the curved background. A direct application may be to use this amplitude to find the probability of the near horizon radiation effects. A simple formulation of such radiation phenomena can be used to find an appropriate quantum mechanical description of the radiation near the black holes horizons.

\subsection{Finding ground state energy via a variational method }
The wave vector $\Psi_{\tau}(x_{+},x_{-})$ in the previous line can not be obtained in closed form. As an attempt to get more physics of the quantum system, we will try find the ground state energy $E_0\leq E_{p,n}$ via a variational method. The method is based on finding the value of a variation parameter $\alpha$ for a trial wave function $\Psi_{\tau}(x_{+},x_{-}|\alpha)$ that minimizes the expectation value of the Hamiltonian: 

	\begin{eqnarray}
	&&<\Psi_{\tau}(x_{+},x_{-}|\alpha)|\hat{H}|\Psi_{\tau}(x_{+},x_{-}|\alpha)>
	\end{eqnarray}
In our case it reduces to minimize the following functional:
\begin{widetext}
	\begin{eqnarray}
&&\bar{k}_{0}^2=Minimize\Big[\frac{\int_{-\infty}^0dy(e^{-y}-1)^{-\bar{r}^2}(F_{\alpha}'(y))^2+\bar{m}^2(e^y\coth^2y)(e^{-y}-1)^{-1-\bar{r}^2}
	F^2_{\alpha}(y)}{\int_{-\infty}^0dy(e^{-y}-1)^{-\bar{r}^2}F^2_{\alpha}(y)}
\Big]\label{k0}
\end{eqnarray}
\end{widetext}
where the trial function $F_{\alpha}(y)$ should satisfy the essential boundary condition presented in the equation (\ref{bc}) for the general wave function $\Psi_{\eta}(y,\zeta)$: 

	\begin{eqnarray}
	&&F_{\alpha}(y)=
	\left\{
	\begin{array}{ll}
	0  & \mbox{if } y\to 0 \\
	0
	&  \mbox{if } y\to-\infty
	\end{array}
	\right.
	\end{eqnarray}

An unnormalized  trial function  on the interval $y\in(-\infty,0)$ with an unweighted scalar product is $F_{\alpha}(y)=\alpha ye^{y}$ because it satisfies the boundary condition given in equation (\ref{bc}), can be used to minimize the functional Eq.(\ref{k0}). We evaluated the integral in Eq. (\ref{k0}) and plugging the result in Eq. (\ref{E}) we obtain for the ground state energy:
\begin{widetext}
\begin{eqnarray}
&&E_{p_{UV},0}\approx\frac{(\hbar\bar{r}\lambda)^2}{mM}\frac{ \left(H_{\bar{r}^2+1}+\gamma -1\right) \psi ^{(0)}\left(\bar{r}^2+2\right)+ \psi
	^{(1)}\left(\bar{r}^2+2\right)-\frac{\pi ^2}{6}+ (\gamma -1) \gamma +\frac{3}{2}}{ \left(H_{\bar{r}^2+1}+\gamma
	-3\right) \psi ^{(0)}\left(\bar{r}^2+2\right)+ \psi ^{(1)}\left(\bar{r}^2+2\right)-\frac{\pi ^2}{6}+
	(\gamma -3) \gamma +\frac{7}{2}}-\frac{\lambda^2p_{UV}^2}{2mM}\label{E0}
\end{eqnarray}
\end{widetext}
here $p_{UV}$ is the ultraviolet cutoff and take it to be a momentum,the computations performed up to $\mathcal{O}(\bar{m}^2)$,  $\gamma$ is the Euler-Mascheroni constant  defined as
\begin{eqnarray}
&&\gamma=\lim_{n\to\infty}\Big(\sum_{k=1}^n\frac{1}{k}-\ln n\Big)
\end{eqnarray}

 with numerical value $\gamma\approx0.577216$. $\psi ^{(n)}(x)$ denotes the PolyGamma function and gives  as the digamma function given as:
 \begin{eqnarray}
 &&\psi ^{(n)}(x)=\frac{d}{dx}\ln\Gamma(x)
 \end{eqnarray}

 Define  $H_{n }=\sum_{k=1}^n\frac{1}{k}$. The harmonic number gives the $n^{\mbox{th}}$ harmonic number.  We ignored the effect of mass $\bar{m}^2$, and the result is represented as the first order approximation. Note that in the semi classical limit where $\hbar\to 0$, the second term in the bracket becomes constant and the first term gives the first order $\mathcal{O}(\hbar^2)$ correction to $E_0$.  
\section{Mode decomposition for Stueckelberg-Schr\"{o}dinger equation  on five dimensional blackhole spacetime}
Blackholes in higher dimensional spacetimes have been invesigated as direct examples for violation of uniquness of the blackholes theorem in four dimensions (see \cite{Mazur:2000pn} for a brief review). In  five dimensional Riemanninan manfolds it is possible to find a doubly spinning black hole with a pair of angular momenta as a five dimensional analog of the axially stationary exterior Kerr metric in four dimensions \cite{Emparan}. Not only blackholes but also black rings and strings have been investigated with different topologies for the horizon in higher dimensional GR and many interesting properties  studied , for example by describing thermodynamics of such higher dimensional black objects using differential geometry \cite{Bravetti:2012hd} 
(see \cite{Emparan} for a comprehensive review on higher dimensional black objects ). If we focus on asymptotically anti-de Sitter black holes
 \cite{examples}, the presence of the cosmological constant  strongly changes  the black hole properties ,for example, making them  unstable \cite{unstable}.\par
 In comparison to the uncharged wave equation solved in the previous section for a 2D toy blackhole, we now study a possible mode decomposition for $\Psi_{\tau}$ in a general five dimensional neutral static-spherically symmetric metric in the presence of the scalar field of the potential model. $V(r)$ is now replaced by the generally $\tau$  dependent function $V_5(\tau,x^{\sigma})$. The metric is given as \cite{d-dim-bh}
 \begin{eqnarray}
&&ds^2=-fdt^2+f^{-1}dr^2+r^2d\Omega_3
 \end{eqnarray}
where $d\Omega_3$ is the line element on the $3$-dimensional unit sphere metric \cite{Frye:2012jj}:
 \begin{eqnarray}
&&d\Omega_3=d\theta_2^2+\sin^2\theta_2d\theta_1^2+\sin^2\theta_2\sin^2\theta_1d\varphi^2.
\end{eqnarray}

where $f$ is defined as
 \begin{eqnarray}
&&f=1-\frac{2m}{r^2}
\end{eqnarray}
 The starting point is to write a gauge transformed version of the eigenvalue-eigenfunction for Hamiltonian Eq. (\ref{H1}) using  a relativistic Hamiltonian with gauge invariant form:

\begin{widetext}
\begin{eqnarray}
&&i\hbar\frac{\partial}{\partial\tau}\Psi_{\tau}(x^\mu)=\Big[\frac{1}{2M\sqrt{-g}}(p_{\mu}-ea_{\mu}(\tau,x^{\sigma}))g^{\mu\nu}(p_{\nu}-ea_{\nu}(\tau,x^{\sigma}))-eV_5(\tau,x^{\sigma})\Big]\Psi_{\tau}(x^\mu)\label{waveeq}
\end{eqnarray}
\end{widetext}

Here the $a_{\mu}$,  may depend on the affine parameter $\tau$ as well as coordinate $x^{\sigma}$,$e$ is charge and the $V_5(\tau,x^{\sigma})$ is the  gauge invariance of the $\tau$ derivative in the canonical quantum mechanical Stueckelberg-Schrodinger equation. It was demonstrated that there exists a conformal transformation of the Hamiltonian which  maps the system at one classical level to
 another conformal metric structure where   the new metric
   is cast into  a Kaluza-Klein effective metric \cite{Horwitz:2009nm}. Using this effective conformal dual metric it was demonstrated that the conformally modified metric emerges from the Hamilton equations \cite{com2}.
 For the sake of simplicity and preserving spherical symmetry  we assume that $V_5(\tau,x^{\sigma})=V(r)$ and
 $a_{\mu}(\tau,x^{\sigma})=A(r)\delta_{\mu}^{\tau}$ and use a seperation of variables as 
 \begin{widetext}
 \begin{eqnarray}
&&\Psi_{\tau}(t,r,\Omega_3)=\sum_{n=0}^{\infty}\int d\Omega_3\int_{-\infty}^{\infty} d\omega e^{-\frac{i}{\hbar}(E_{n}\tau+\hbar\omega t)}Y_n(\Omega_3)R_n(r,\omega)
 \end{eqnarray}
\end{widetext}
with the above wave function in Eq. (\ref{waveeq}), we can write the radial differential equation for $R_n(r,\omega)$  as
 \begin{widetext}
 	\begin{eqnarray} 	
 	\label{radial-R}&&R''_n(r,\omega)+\Big(2i\hbar e+\frac{2}{r}-\frac{f'}{f}
 	\Big)R'_n(r,\omega)\\&&\nonumber+\Big[(\omega f)^2+\frac{n(n+2)f}{r^2}+i\hbar e(A'-\frac{A(r)f'}{f})+e^2A(r)^2+eV(r)f-\frac{2ME f}{\hbar^2}
 	\Big]R_n(r,\omega)=0.
 	\end{eqnarray}
 \end{widetext}
 
 where we introduce the spherical harmonics $Y_n(\Omega_3)$ as eigenfunctions of the Laplacian operator on $\Omega_3$ satisfying  the following eigenvalue-eigenfunction 
linear equation \cite{Frye:2012jj}:
\begin{eqnarray}
&&\Delta_3Y_n(\Omega_3)=-n(n+2)Y_n(\Omega_3)
\end{eqnarray}
We can find approximate solutions  for the radial equation (\ref{radial-R}) near horizon $f(r_h)=0$ and at infinity $r\to\infty$ if we suppose that $A$ near the horizon 
, remains finite; calling it $A_h$, it vanishes at infinity as well as $V(r)$.
\begin{widetext}
	\begin{eqnarray}
	&&\{A(r),V(r)\}=
	\left\{
	\begin{array}{ll}
\{A_h,V_h\}	  & \mbox{if } r\to r_h \\
	0
	&  \mbox{if } r\to\infty
	\end{array}
	\right.
	\end{eqnarray}
\end{widetext}
The approximated solutions are represented  as follows:
\begin{widetext}
	\begin{eqnarray}
&&R_n(r,\omega)=
\left\{
\begin{array}{ll}
\frac{A_h e \hbar(r-r_h)}{\sqrt{2}}\Big( a_n I_2\left(2 \sqrt{i A_h e \hbar(r-r_h)}\right)+b_n K_2\left(2 \sqrt{i A_h e \hbar
		(r-r_h)}\right)\Big) & \mbox{if } r\to r_h \\
e^{-ie\hbar r}\Big(c_ne^{-kr}+d_n e^{kr}
\Big)
&  \mbox{if } r\to\infty
\end{array}
\right.
\end{eqnarray}
\end{widetext}
where $k=\sqrt{\frac{2ME}{\hbar^2}-\omega^2-(e\hbar)^2-iA'_{\infty}e\hbar}$ and $I_n(x),K_n(x)$ give the modified Bessel function of the first and second  kinds. Note that $R_n(r_h,\omega)=\frac{i b_n}{2 \sqrt{2}}$ remains finite. The asymptotic radial  eigenfunctions can be used to find asymptptotic Green functions for Eq. (\ref{waveeq}). It will be interesting if we can use this radial solution to find the most probable distance of the test particle from the horizon $r_h$. 

\section{Summary }
 There are still several attempts to obtain  a unified theory in which gravity as a classical gauge theory can be naturally quantized in the canonical way (as well as path integral method). It is clear for me that  any such consistant quantum gravity should be initiated from same steps as in the mathematical formulation of the 
  quantum theory, passing through classical Poisson brackets to Dirac's communtators and then studying the time evolution equation both for the state vector (in Schr\"{o}dinger picture ) and quantum operator (in Heisenberg approach ). A simple and physically understandable approach  where gravity is quantized canonically was recently  proposed by  Horwitz. In our letter  we present exact solutions for classical and quantum dynamics of GR in this canonical method. As a curved background I focused on a two dimensional toy model blackhole as an integrable system.  A classical geodesic equation wih corrections was explicitly derived and integrated for a test particle in a simple radial potential field. Using the  classical solution obtained in my letter, we can support the idea of classical fields as effective fields. By investigating the linear perturbations of the metric , one is able to explain the early (or pre-early)
   inflation and thermal spectrum of the matter contents. Furthermore, this solution can be used to build the appropriate forms for the Green function. It will provide the  chronological time propagators and using those propagators we can study the causal structeres of the spacetime metric at early universe. 
  In the quantum mechanical version of this canonical theory, we investigated exact solutions
   for the modified Schr\"{o}dinger equation on the curved background. 
The complete set of the exact mode solutions were calculated  for the modified wave equation as well as an estimated ground state energy level. The quantum mechanical amplitudes  can be used to  explain the high energy excitations in the vicinity of the curved background. A direct application will be to  find the probability of near horizon radiation effectsand to justify the origin  of the radiation near the black holes horizons. 
As an attempt to extend the theory to electromagnetism we solved the wave equation for a test charge particle. Asymptotic solutions obtained can be used to construct Green functions and causal propagators. 
  This work will be  continued in a forthcoming paper on Green functions and relevant field theoretical aspects.

\section{Acknowledgment} 
I  thank Prof. Lawrence P. Horwitz for carefully reading my first draft, very useful comments, corrections and discussions. I thank the anonymous referees for intuitive comments and thorough criticism on my manuscript. 
I would like to acknowledge the support of Sultan Qaboos University under the Internal Grant  (IG/SCI/PHYS/19/02). 



\end{document}